\documentstyle[emulateapj]{article}

\lefthead{GOODS-N 70$\mu$\lowercase{m} Source Counts}
\righthead{Frayer et al.}
\slugcomment{Accepted ApJL, 2006 June 26}

\newcommand{\lsun}{\,L_{\odot}}
\newcommand{\nW}{\,\rm{nW~m}^{-2}{\rm sr}^{-1}}

\begin{document}

\title{{\it Spitzer} 70$\mu$\lowercase{m} Source Counts in
GOODS-North}

\author{D.\ T.\ Frayer\altaffilmark{1},
M.\ T.\ Huynh\altaffilmark{1},
R.\ Chary\altaffilmark{1},
M.\ Dickinson\altaffilmark{2},
D.\ Elbaz\altaffilmark{3},
D.\ Fadda\altaffilmark{1},
J.\ A.\ Surace\altaffilmark{1},
H.\ I.\ Teplitz\altaffilmark{1},
L.\ Yan\altaffilmark{1},
B.\ Mobasher\altaffilmark{4}
}

\altaffiltext{1}{{\it Spitzer} Science Center, California Institute of
Technology 220--06, Pasadena, CA 91125; frayer@ipac.caltech.edu.}

\altaffiltext{2}{National Optical Astronomy Observatories, 950 North
Cherry Avenue, Tucson, AZ 85726.}

\altaffiltext{3}{DSM/DAPNIA/Service d'Astrophysique, CEA/SACLAY, 91191
Gif-sur-Yvette Cedex, France.}

\altaffiltext{4}{Space Telescope Science Institute, 3700 San Martin
Drive, Baltimore MD 21218.}

\begin{abstract}

We present ultra-deep {\it Spitzer} 70$\mu$m observations of
GOODS-North (Great Observatories Origins Deep Survey).  For the first
time, the turn-over in the 70$\mu$m Euclidean-normalized differential
source counts is observed.  We derive source counts down to a flux
density of 1.2\,mJy.  From the measured source counts and fluctuation
analysis, we estimate a power-law approximation of the faint 70$\mu$m
source counts of $dN/dS \propto S^{-1.6}$, consistent with that
observed for the faint 24$\mu$m sources.  An extrapolation of the
70$\mu$m source counts to zero flux density implies a total
extragalactic background light (EBL) of $7.4\pm1.9\nW$.  The source
counts above 1.2\,mJy account for about 60\% of the estimated EBL.
From fluctuation analysis, we derive a photometric confusion level of
$\sigma_c = 0.30\pm0.15$\,mJy ($q=5$) for the {\it Spitzer} 70$\mu$m
band.

\end{abstract}


\section{Introduction}

Deep 24$\mu$m observations (Chary et al. 2004; Papovich et al. 2004;
Fadda et al. 2006) have demonstrated the ability of the Multiband
Imaging Photometer for {\it Spitzer} (MIPS, Rieke et al. 2004) to
study the mid-infrared (mid-IR) properties of high-redshift galaxies
(Yan et al. 2004; Le Floc'h et al. 2004, 2005; P\'{e}rez-Gonz\'{a}lez
et al. 2005; Daddi et al. 2005; Papovich et al. 2006; Caputi et
al. 2006).  The interpretation of the 24$\mu$m data are complicated by
the presence of strong emission and absorption features (e.g., Armus
et al. 2004) redshifted into the 24$\mu$m band.  Observations at
longer wavelengths, such as 70$\mu$m which is closer to the
far-infrared peak of the spectral energy distribution (SED) and is
away from the strong mid-IR features, are crucial for constraining the
infrared luminosities and star-formation rates.

The previous deep Guaranteed Time Observer (GTO) surveys did not
achieve sufficient sensitivity at 70$\mu$m to detect distant luminous
infrared galaxies (LIRGs; $10^{11}\lsun \la L_{ir} \la 10^{12}\lsun$),
without stacking 70$\mu$m data for a large number of 24$\mu$m-selected
sources (Dole et al. 2006).  Much deeper observations are needed at
70$\mu$m to individually detect the $z\sim1$ LIRGs that account for
the majority of the extragalactic background light (Elbaz et al. 2002;
Lagache et al. 2004).  In this letter, we present initial results for
the deepest 70$\mu$m survey taken to date with {\it Spitzer}.

\section{Observations}

The ultra-deep 70$\mu$m observations of the northern field of the
Great Observatories Origins Deep Survey (GOODS-N) were carried out in
Cycle-1 of the General Observer (GO) program ({\it Spitzer} program
3325).  The survey covers the central $10^{\prime}\times 10^{\prime}$
of GOODS-N to a depth of 10.6\,ks.  The data were taken using
small-field photometry mode with 10\,MIPS-second data collection
events (DCEs).  The field was observed with an 8 position cluster map
for each astronomical observational request (AOR).  The observations
were repeated with 12 AORs taking 34.5 hours of observatory time in
total.  The mapping order and dither positions of the cluster
positions within the AORs were varied to provide uniform coverage and
data quality across the field.  The data were embargoed until after
the GTO proprietary period and were released to our team in 2005
August.  In addition to the GO data, we used the MIPS GTO data of
GOODS-N ({\it Spitzer} program 81, Dole et al. 2004a).  The GTO data
were taken in slow scan mode with one degree scan legs and have an
integration time of 600\,s at 70$\mu$m, covering an area of
0.6\,deg$^2$.

\section{Data Reduction}

The raw data were downloaded from the {\it Spitzer} Science Center
(SSC) archive and were processed from scratch using the offline
Germanium Reprocessing Tools (GeRT, S13 version 1.0).  The
instrumental artifacts in the basic calibration data (BCDs) were
removed adopting the filtering techniques used for the reduction of
the extragalactic First Look Survey (xFLS, Frayer et al. 2006).  The
BCD pipeline processing and filtering procedures were optimized for
these deep photometry data.  We adopted the updated S13 calibration,
which assumes an absolute flux calibration factor based on stellar
SEDs of 702\,MJy\,sr$^{-1}$ per MIPS-70 data unit.  We then multiplied
the data by the color correction factor of 1.09 to place the data on a
constant $\nu f_\nu$ scale, which is also the appropriate color
correction (within 2\%) for a wide range of possible galaxy SEDs (see
SSC web pages for calibration and color-correction details).  In
comparison, the calibration correction adopted here is 3.4\% larger
than the calibration adopted for the xFLS analysis (Frayer et
al. 2006).

The filtering of the data is a crucial aspect in the processing.  For
the 70$\mu$m photometry mode, calibration stimulator (stim) flashes
are used every 6 DCEs, and latents due to these stim flashes
accumulate over time.  To remove stim flash latents and additional
artifacts, we used a median column filter followed by a median
high-pass time filter per pixel (with a filter width of 16 DCEs).  The
positions of bright sources in the BCDs were flagged so that the
filtering corrections were not biased by the presence of sources.  The
median filtering techniques yield small offsets from zero in the
average level of the filtered-BCDs (fBCDs).  These offsets correlate
with the DCE position within the stim cycle and were removed by
subtracting the median level from each fBCD.

The data were coadded onto a sky grid with $4\arcsec$ pixels using the
SSC mosaicing and source extraction software (MOPEX, version 112505).
Sources were extracted using MOPEX point source response function
(PRF) fitting.  In crowded regions, the 24$\mu$m positions (R. Chary
et al., in preparation) were used for deblending.  For optimal source
extraction, an accurate uncertainty image is needed.  The uncertainty
image was constructed by combining the noise per pixel based on
repeated observations with the local spatial pixel-to-pixel dispersion
after the extraction of bright sources.  The average level of the
uncertainty image was then scaled to match the average empirical
point-source noise derived by making multiple aperture measurements at
random locations throughout the residual mosaic after source
extraction.  The scale factor between the aperture derived point
source noise (including the aperture correction) and the pixel surface
brightness noise is $\sigma$(point source)/$\sigma(4\arcsec$ pixels)
$= 10.9\pm1.1$\, [mJy/(MJy\,sr$^{-1}$)].  The average point source
noise for the ultra-deep area (after the extraction of sources) is
0.53\,mJy ($1\sigma$).  The applicability of the uncertainty image for
point source extraction was verified by obtaining PRF fits with
$\chi^2\simeq 1$ for sources throughout the image.

\section{Results}

\subsection{Estimate of Confusion}

Previous surveys with {\it Spitzer} are not deep enough to measure the
photometric confusion noise at 70$\mu$m, and the estimated confusion
level is based on the observed bright source counts and models of
galaxy evolution (Dole et al. 2004b).  With the ultra-deep data of
GOODS-N, we can directly measure the confusion level at 70$\mu$m.  We
define the instrument noise (including photon noise, detector noise,
and noise associated with the data processing) as $\sigma_I$.  The
total noise ($\sigma_{T}$) represents the noise after the extraction
of sources above a limiting flux density ($S_{lim}$), and the
confusion noise ($\sigma_{c}$) represents fluctuations due to sources
with flux densities below $S_{lim}$.  As defined here, $\sigma_c$ is
the ``photometric'' confusion noise, following the terminology of
Dole, Lagache, \& Puget (2003).  In the direction of GOODS-N, the
contribution of Galactic cirrus to the confusion noise is negligible
($\sim 0.01$--0.02\,mJy), based on the relationship given by Dole et
al. (2003) and the calculations of Jeong et al. (2005).

The instrument noise was estimated empirically by subtracting pairs of
data with the same integration time and covering the exact same region
on the sky to remove sources and any remaining residuals from the sky
after filtering.  The measured instrument noise integrates down nearly
with $t^{-0.5}$ (Fig. 1).  For these data, we find $\sigma_{I}^{2}
\propto t^{-1}(1 + \beta t^{0.5})$, where $\beta=0.04$ for integration
time $t$ in units of ks.  The functional form of this relationship is
based on empirical results from several different data sets, and the
$\beta$ parameter depends on the background level and the quality of
the data reduction.  We use the above function to extrapolate the
instrument noise from half the data to the full data set and derive
$\sigma_{I} = 0.0399\pm 0.0036$\,MJy\,sr$^{-1}$.

Since the total noise image (after source extraction) and instrumental
noise image have nearly Gaussian distributions, the confusion noise
can also be approximated by a Gaussian and is given by $\sigma_c=
(\sigma_T^2-\sigma_I^2)^{0.5}$.  We iterate between source extraction
at different limiting flux densities and confusion noise measurements
until we converge to a solution with $q\equiv S_{\lim}/\sigma_c =5$.
For the $q=5$ solution, we derive $\sigma_{T} = 0.0485\pm
0.0034$\,MJy\,sr$^{-1}$ and $\sigma_{c} = 0.0276\pm
0.0079$\,MJy\,sr$^{-1}$, for a limiting source flux density of
S70$=1.5$\,mJy.\footnote{S70$=$S$_\nu$(71.4$\mu$m) and
S24$=$S$_\nu$(23.7$\mu$m) throughout this paper.}  Including the
additional systematic uncertainties of the absolute calibration scale
(10\%) and the conversion between point source noise and surface
brightness noise (10\%, \S3), we derive a point source confusion noise
of $\sigma_{c} = 0.30\pm 0.15$\,mJy ($q=5$).

In comparison, the predictions of Dole et al. (2003, 2004b) suggest a
$q=5$ photometric confusion level of $\sigma_c\simeq 0.28$\,mJy,
depending on the exact shape of the differential source counts.  The
measured confusion level of 0.3\,mJy agrees well with the predicted
photometric confusion level.  However, the source density criterion
(SDC) confusion limit of 3.2\,mJy adopted by Dole et al. is more than
a factor of two higher than the limiting flux density derived here.
The Dole et al. SDC limit corresponds to $q\simeq7$ and a high
completeness level of $>90$\%.  Sources can be extracted at lower
completeness levels, and counts can be derived reliably well below
this SDC limit by making use of the 24$\mu$m data to help extract the
faintest sources.

\subsection{Source Counts}

The source counts were derived separately for the central
$10^{\prime}\times10^{\prime}$ ultra-deep field, for the wide
0.614\,deg$^{2}$ GTO deep field, and for the
$12\farcm9\times12\farcm9$ intermediate field which includes the
ultra-deep field and the surrounding regions of intermediate depth
between the GTO and ultra-deep surveys (Table~1).  Analysis was done
on the ``intermediate'' field for better statistics at S70$>4$\,mJy.
The central ultra-deep area has a deficiency ($\sim 30$--50\%) in
number of sources with flux densities between 5--10\,mJy in
comparison to typical areas surrounding GOODS-N, presumably due to
cosmic variance.  The intermediate and GTO fields allow the
measurement of source counts for flux densities (4--12\,mJy) not well
sampled by the ultra-deep GOODS-N and the shallow xFLS surveys.  The
combination of the GOODS-N and the xFLS data sets yields source counts
over the flux density range from 1.2\,mJy to 455\,mJy (Fig. 2).  In
comparison, the counts presented here are 12.5 times deeper than the
counts previously published by the MIPS Team (Dole et al. 2004a) and
about 7 times deeper than the counts based on the xFLS verification
field (Frayer et al. 2006).

Sources were extracted for signal-to-noise (S/N) ratios of S/N$>$2.5
for the ultra-deep field and S/N$>4$ for the intermediate and GTO-deep
fields.  The uncertainty image (\S3) was used to represent the
point-source noise as a function of position.  The number of spurious
sources for each flux bin was estimated by performing source
extraction on the negative image.  Reliability degrades significantly
below S/N$<4$.  To compensate for spurious sources at low S/N, we
required the presence of 24$\mu$m counterparts in the ultra-deep
field.  Since the GOODS 24$\mu$m data (M. Dickinson et al., in
preparation) are about 100 times more sensitive in flux density, very
few, if any, 70$\mu$m-only detections are expected.  Within the
ultra-deep field, the chance coincidence within 8$\arcsec$ ($\simeq 2$
times the positional rms uncertainty of the faintest 70$\mu$m sources)
of a 24$\mu$m counterpart with S24$>80\mu$Jy is 20\%.  By requiring
24$\mu$m counterparts, we removed 80\% of the spurious sources, and
the observed counts were then corrected assuming a 20\% chance match
for the remaining spurious sources.

After correcting for reliability, the observed source counts are
corrected for completeness and flux biasing (i.e., the scattering of
faint sources into brighter flux density bins) using the Monte Carlo
approach described by Chary et al. (2004).  Simulated point sources
spanning the full range of flux densities were injected at random
positions into each image (one at a time with 3$\times10^4$
repetitions per field).  By using the simulations and the same source
extraction methods, we populate a $P_{ij}$ matrix representing the
probability that a galaxy with input flux density $i$ is detected and
measured with output flux density $j$.  By summing over the elements
of the $P_{ij}$ matrix, the true counts are derived from the observed
counts.  This Monte Carlo approach allows the derivation of the counts
at low completeness levels; the completeness level for the faintest
flux bin (1.2--1.6\,mJy) is only 30\% (Table 1).  The uncertainties on
the counts include the uncertainties of the reliability corrections,
the Poissonian errors propagated through the $P_{ij}$ matrix, and the
10\% absolute flux calibration uncertainty combined in quadrature.

The Euclidean-normalized differential source counts turn over around
8--10\,mJy (Fig. 2).  The observed counts are consistent within 30\%
of the Lagache et al. (2004) model for bright flux densities.  We find
a slightly larger number of faint galaxies (S70$<$3\,mJy) than
predicted by Lagache et al. (2004), which may be due to cosmic
variance in the direction of GOODS-N.  Ultra-deep 70$\mu$m observations
over larger areas and along different lines of sight are needed to
constrain the models more accurately at faint flux densities.

At low flux densities, a weighted least-squares fit to the
differential source counts yields $dN/dS \propto S^{-\alpha}$ with
$\alpha=1.6\pm0.6$.  This is consistent with the power law of
$\alpha=1.6\pm0.1$ derived for the faint 24$\mu$m sources (Chary et
al. 2004).  The observed fluctuations also constrain the faint source
counts.  We estimate the extrapolated source counts down to the
confusion limit via simulations.  Different populations of sources
covering a wide range of $\alpha$ values and normalizations at 2\,mJy
were randomly injected into the instrumental noise image and then
extracted using the same techniques carried out for the confusion
measurement.  The best-fit solution for simulations consistent with
the observed constraints is $\alpha=1.63\pm0.34$.  At the confusion
level of 0.3\,mJy, we derive an extrapolated value for the
Euclidean-normalized differential source counts of
$(dN/dS)S^{2.5}=290\pm200$\,gal\,sr$^{-1}$\,Jy$^{1.5}$.

\subsection{Extragalactic Background Light}

The expected extragalactic background light (EBL) due to infrared
galaxies at 70$\mu$m is fairly uncertain due to the difficulty of
deriving accurate zodiacal light corrections for the Diffuse Infrared
Background Experiment (DIRBE) measurements (e.g., Wright 2004). By
interpolating between the 24$\mu$m EBL value (Papovich et al. 2004),
the 60$\mu$m value (Miville-Dech\^{e}nes, Lagache, \& Puget 2002), and
the DIRBE 100$\mu$m and 140$\mu$m measurements (Wright 2004), we
estimate a predicted EBL level of $\nu I_{\nu}=10\pm5\nW$ at
71.4$\mu$m.  In comparison, Dole et al. (2006) estimates a total EBL
of $7.1\pm1.0\nW$ from 70$\mu$m stacking analysis and the extrapolation
of the 24$\mu$m counts, and the Lagache et al. (2004) model predicts a
value of $6.4\nW$.

By summing over the observed source counts for S70$>1.2$\,mJy
(including a small contribution from sources brighter than 455\,mJy
based on the Lagache et al. 2004 model), we derive a contribution of
$4.3\pm0.7\nW$ to the EBL.  This is about 60\% of the total EBL.  By
extrapolating ($\alpha=1.63\pm0.34$) the 70$\mu$m source counts down
to the confusion level, we derive a contribution of $5.5\pm1.1\nW$ for
S70$>0.3$\,mJy, and the extrapolation to zero flux density yields an
estimated total EBL of $7.4\pm1.9\nW$.  The uncertainties on the EBL
measurements include an additional 10\% systematic uncertainty to the
error budget, accounting for the uncertainties associated with the
absolute calibration and color corrections.

\section{Concluding Remarks}

Based on ultra-deep 70$\mu$m observations, we derive source counts
down to a flux density of 1.2\,mJy, directly resolving about 60\% of
the EBL.  The total fraction of the EBL estimated for sources down to
the confusion level ($\sigma_c\simeq0.3$\,mJy, $q=5$) is about 75\%.
A power-law extrapolation to zero flux density implies a total EBL of
$7.4\pm1.9\nW$ at 71.4$\mu$m.  This is consistent with the value
predicted based on EBL measurements at other wavelengths, the value
predicted from the Lagache et al. (2004) model, and the value derived
from the extrapolation of the 24$\mu$m counts and 70$\mu$m stacking
analysis (Dole et al. 2006).  However, the uncertainties on the
results leave open the possibility of a significant population of
sources at low 70$\mu$m flux densities that are not accounted for in
the models, such as highly obscured $z\sim 1$ AGNs as proposed to
account for the hard X-ray background (e.g., Worsley et al. 2005).
Studies of the counterparts of the faint 70$\mu$m population are
ongoing and will help to constrain the infrared luminosities and the
relative fraction of AGN versus starburst-dominated galaxies in the
high-redshift {\it Spitzer}-selected surveys.

We thank our colleagues associated with the {\it Spitzer} mission who
have made these observations possible.  This work is based on
observations made with the {\it Spitzer Space Telescope}, which is
operated by the Jet Propulsion Laboratory, California Institute of
Technology under NASA contract 1407.


\begin{figure}
\plotone{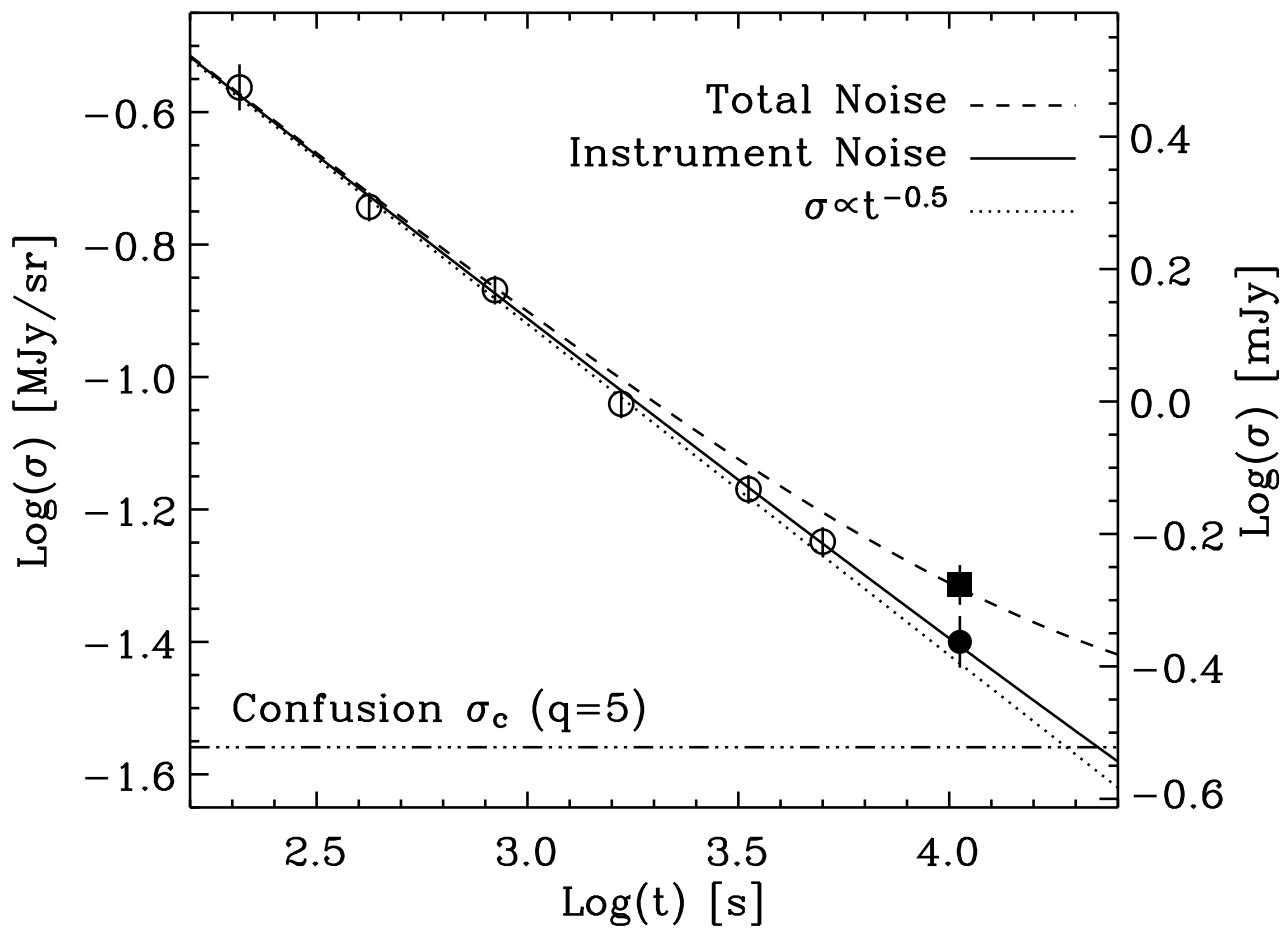}
\caption{Noise for $4\arcsec$ pixels as a function of integration time
($1\sigma$).  The corresponding point-source noise in mJy is shown at
the right.  Instrumental noise measurements are shown as open circles
and are represented by the solid line.  For comparison, the dotted
line shows a $t^{-0.5}$ function.  The derived confusion level
($\sigma_{c}$) is shown by the dashed-dotted line, and the total noise
after the extraction of sources with $S70>5\sigma_{c}$ is shown by the
dashed line.  The total noise and instrument noise for the ultra-deep
field are shown by the solid square and solid circle, respectively.}
\end{figure}

\begin{figure}
\plotone{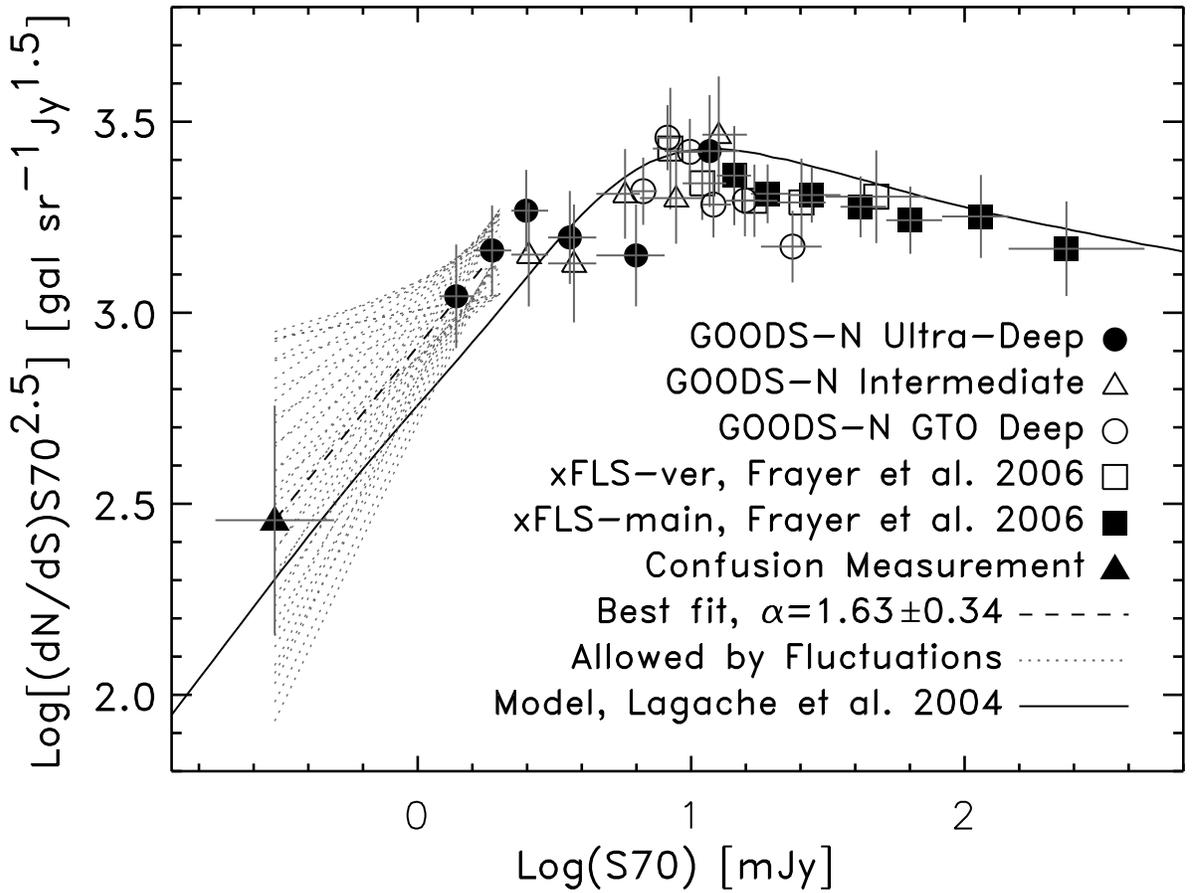}
\caption{The 70$\mu$m differential source counts from the ultra-deep
GOODS-N field are shown as filled circles, the intermediate field
counts are shown as open triangles, while the results from the GTO
deep field are shown as open circles.  The xFLS number counts (Frayer
et al. 2006), corrected for the updated calibration, are shown as
squares.  The vertical bars represent the total errors, and the
horizontal line segments show the sizes of each bin (Table~1).  The
model of Lagache et al.(2004) is shown by the solid line.  The
dashed-line represents the best fit extrapolation of the counts
($dN/dS \propto S^{-\alpha}$) down to the confusion level shown by the
solid triangle, and the allowable range of extrapolated counts
constrained by fluctuation analysis is given by the dotted-lines.}
\end{figure}


\begin{deluxetable}{cccccccc}
\tablecaption{GOODS-N 70$\mu\lowercase{m}$ Source Counts}
\tablehead{
\colhead{S(low)}&
\colhead{S(high)} &
\colhead{S(avg)} &
\colhead{Observed}&
\colhead{Reliability}&
\colhead{Completeness}&
\colhead{Corrected}&
\colhead{Log[$(dN/dS)\,S^{2.5}$]}\\
\colhead{[mJy]}&
\colhead{[mJy]}&
\colhead{[mJy]}&
\colhead{Number}&
\colhead{}&
\colhead{}&
\colhead{Number}&
\colhead{[gal sr$^{-1}$ Jy$^{1.5}$]}
}
\startdata

 1.2&  1.6&   1.38&   21& 0.75& 0.30& 52.4$\pm$14.3&  3.043$\pm$0.136\\
 1.6&  2.2&   1.87&   28& 0.85& 0.49& 48.7$\pm$11.0&  3.163$\pm$0.118\\
 2.2&  3.0&   2.49&   31& 0.97& 0.75& 40.2$\pm$ 7.8&  3.267$\pm$0.107\\
 3.0&  4.5&   3.60&   21& 0.97& 0.80& 25.6$\pm$ 6.1&  3.197$\pm$0.122\\
 4.5&  8.0&   6.29&   14&  1.0& 1.05& 13.3$\pm$ 3.6&  3.150$\pm$0.133\\
 8.0& 15.0&  11.68&   11&  1.0& 1.04& 10.6$\pm$ 3.2&  3.423$\pm$0.147\\
\ \\
 2.2&  3.0&   2.55&   28& 0.75& 0.43& 48.7$\pm$13.4&  3.152$\pm$0.136\\
 3.0&  4.5&   3.73&   26& 0.69& 0.54& 33.5$\pm$10.8&  3.129$\pm$0.154\\
 4.5&  6.5&   5.73&   20&  1.0& 0.86& 23.2$\pm$ 5.2&  3.311$\pm$0.117\\
 6.5& 11.0&   8.81&   19&  1.0& 1.10& 17.3$\pm$ 4.0&  3.300$\pm$0.119\\
11.0& 16.0&  12.62&   10&  1.0& 0.87& 11.5$\pm$ 3.7&  3.466$\pm$0.153\\
\ \\
 6.0&  7.5&   6.68&  101&  0.86& 0.54& 160.1$\pm$21.9&  3.318$\pm$0.088\\
 7.5&  9.0&   8.19&   96&  0.95& 0.69& 132.5$\pm$16.8&  3.458$\pm$0.085\\
 9.0& 11.0&   9.89&   73&  0.99& 0.71& 101.4$\pm$13.2&  3.421$\pm$0.086\\
11.0& 14.0&  12.07&   62&  1.0 & 0.92&  67.3$\pm$ 8.7&  3.283$\pm$0.086\\
14.0& 18.0&  15.72&   45&  1.0 & 0.95&  47.5$\pm$ 7.4&  3.294$\pm$0.094\\
18.0& 30.0&  23.48&   43&  1.0 & 1.09&  39.6$\pm$ 6.1&  3.173$\pm$0.094

\enddata

\tablecomments{The first 6 rows are for the ultra-deep field
  (0.0277\,deg$^{2}$), the next 5 rows are for the intermediate field
  (0.0462\,deg$^{2}$), and the remaining rows are for GTO-deep field
  (0.614\,deg$^{2}$). The corrected counts are equal to the observed
  counts multiplied by the reliability value and divided by the
  completeness correction.  The uncertainties for the
  Euclidean-normalized differential counts [$(dN/dS)\,S^{2.5}$]
  include the Poissonian noise, the uncertainties associated with the
  reliability, completeness, and flux biasing corrections, and the
  absolute calibration uncertainty of 10\%.}

\end{deluxetable}


\begin{references}

\reference{} Armus, L., et al. 2004, ApJS, 154, 178

\reference{} Caputi, K. I., et al. 2006, ApJ, 637, 727

\reference{} Chary, R., et al. 2004, ApJS, 154, 80

\reference{} Daddi, E., et al. 2005, ApJ, 631, L13

\reference{} Dole, H., Lagache, G., \& Puget, J.-L. 2003, ApJ, 585,
617

\reference{} Dole, H., et al. 2004a, ApJS, 154, 87

\reference{} Dole, H., et al. 2004b, ApJS, 154, 93

\reference{} Dole, H., et al. 2006, A\&A, 451, 417

\reference{} Elbaz, D., Cesarsky, C. J., Chanial, P., Aussel, H.,
Franceschini, A., Fadda, D., \& Chary, R. R. 2002, A\&A, 384, 848

\reference{} Fadda, D., et al. 2006, AJ, 131, 2859

\reference{} Frayer, D. T., et al. 2006, AJ, 131, 250

\reference{} Jeong, W.-S., Lee, H. M., Pak, S., Nakagawa, T., Kwon,
S. M., Pearson, C. P., \& White, G. J. 2005, MNRAS, 357, 535

\reference{} Lagache, G., et al. 2004, ApJS, 154, 112

\reference{} Le Floc'h, E., et al. 2004, ApJS, 154, 170

\reference{} Le Floc'h, E., et al. 2005, ApJ, 632, 

\reference{} Miville-Desch\^{e}nes, M.-A., Lagache, G., \& Puget,
J.-L. 2002, A\&A, 393, 749

\reference{} Papovich, C., et al. 2004, ApJS, 154, 70

\reference{} Papovich, C., et al. 2006, ApJ, 640, 92

\reference{} P\'{e}rez-Gonz\'{a}lez, P. G., et al. 2005, ApJ, 630, 82

\reference{} Rieke, G. H., et al. 2004, ApJS, 154, 25

\reference{} Worsley, M. A., et al. 2005, MNRAS, 357, 1281

\reference{} Wright, E. L. 2004, NewAR, 48, 465

\reference{} Yan, L., et al. 2004, ApJS, 154, 60

\end{references}
\end{document}